\documentclass[a4paper,12pt]{article}

\begin{document}
            Particle and Field Symmetries and Noncommutative Geometry

                               Ajay Patwardhan

                    Physics Department, St.Xavier's College,

                    Mahapalika Marg, Mumbai 400001, India.

                     *  e-mail : ajpatwar@imsc.res.in

                      \qquad           Abstract

The development of Noncommutative geometry is creating a reworking and new possibilities in physics. This paper identifies some of the commutation and derivation structures that arise in particle and field interactions and fundamental symmetries. The requirements of coexisting structures, and their consistency, produce a mathematical framework that underlies a fundamental physics theory.

Among other developments in Quantum theory of particles and fields are the symmetries of gauge fields and the Fermi-Bose symmetry of particles. These involve a gauge covariant derivation and the action functionals ; and commutation algebras and Bogoliubov transforms. The non commutative Theta form introduces an additional and fundamental structure . This paper obtains the interrelations of the various structures; and the conditions for the symmetries of Fermionic/Bosonic particles interacting with Yang -Mills gauge fields. Many example physical systems are being solved , and the mathematical formalism is being created to understand the fundamental basis  of physics.

 \qquad  1.Introduction

The mathematical structures of the  physics of particles and fields were developed using commutative and non commutative algebra, and Euclidean and non Euclidean Geometry. This led to Quantum Mechanics and General Relativity,respectively. The Quantum Field theory of Gauge Fields describes all fundamental interactions, including gravity, as holonomy and action integrals. It has succeeded phenomenologically, inspite of some difficulties. Consistency requirements have led to a number of symmetries, including supersymmetry. Loop space quantum gravity and string and brane theories have evolved as a development of quantum theory of interactions. These are also connected to the evolving subject of non commutative geometry.[Ref ]

The dynamical variables in a quantum theory have a commutation algebra. A non commutative structure has been introduced in a wide variety of physics ; with length scales from Planck length in quantum space time, to magnetic length in quantum Hall effect. The new (non)commutation structure introduces a derivation (as a bracket operation), which acts in addition to the Lie and covariant derivatives. In the spacetime  manifold , a discrete topology and a length scale parameter cause changes in the definitions of the metric tensor,Riemann tensor, Ricci tensor and the Einstein equations.Will this be a' correction' to the standard theory. Generally the non commutative structure may not have dimensions of length, but it can have dimensions of whatever dynamical variables are considered; hence it is a general feature. The $\theta$ structure as a commutation and derivation will ,along with the connection form , create a generalised derivative operator. For example,\qquad $\nabla_i =\partial_i -\Gamma_i +\theta{ij}\partial_j ;\qquad (\nabla_i\nabla_j -\nabla_j\nabla_i)V^k = R^k_{ijl} V^l $ is a possible generalisation for the curvature tensor. [Ref ]

  \qquad  2.The Non commutative structure

The classical phase space for a physical system,is equipped with the Poisson Bracket as well as the noncommutative structure.It leads to a quantisation (first and second), in which the linear ,canonical Bogoliubov transforms of the operators , which preserve commutators, get modified. This works for Bosonic (commutators), as well as Fermionic (anticommutators), and 'q' deformed  variables.The requirements of commutation and derivation are, respectively, algebraic and analytic conditions. They act on the geometry of the dynamical variables, that describe the physical system. They are being formulated in the subject of non commutative geometry and its applications to physics.[Ref]

The Moyal bracket and the theta form provide the derivation and commutation in the examples considered in this paper.In a finite quantum physical system of Bosonic and Fermionic variables, the Bogoliubov and non commutative geometry properties give rise to an understanding of the supersymmetric transformation. This could have a realisation in nuclear spectra, in condensed matter  anyons,and in the quark-gluon particle physics. The comparison is made with the quantum Hall effect system, which is well known. In a Yang -Mills gauge field theory, the gauge symmetry and non commutative geometry together give rise to a reworking of the holonomy and action integrals. The modified classical theory and quantum corrections for gravity may require a comparison with the quantum Riemann geometry of spacetime. [Ref ]

 The Structure of noncommutative  $ \theta$ forms in the theory of dynamical variables interacting with Gauge fields requires that, two functions $f,g$ have a product\qquad $ f*g = fg + i/2 \Theta ^{ij} \partial_if\partial_jg$.  This is the expansion of the Moyal bracket\qquad $exp(\theta^{ij}\partial/\partial x^i \partial/\partial x^j )$ to order $ \theta^2$. The antisymmetric product \qquad $f*g -g*f $ defines the $\theta $ form. This leads to the commutator defined on the coordinates as $ [x^i ,x^j] = x^i*x^j - x^j*x^i = i\theta ^{ij} $  . The $ [\partial_i ,x^j] =\delta^{j}_{i} $ and \qquad$ [\partial_i ,\partial_j] =0 $.

  For example, the $\theta^{ij} $ could be $L^2 \epsilon^{ij}$, with a length parameter  L, and an antisymmetric $\epsilon$. In the quantum Hall effect the magnetic length $ L^2 $ varies inversely as the magnetic field. The representation $ [x^i,p^j] =i\delta^{ij} $ leads to $ P^i = -i\delta^{ij}\partial/\partial x^j  $. The derivation is $-i\theta^{ij}\partial/\partial x^j $  with $[x^i,f] =i\theta^{ij}\partial_jf$ .

The presence of both the canonical and the non commutative structure requires , for consistency, that $ [p^i,p^j]= i(\theta^{ij})^{-1}$. The Lie algebra $ [L^i,L^j] = F^{ij}_k L^k $ and the Poisson Bracket structure, $(\epsilon_{ij} \partial/\partial x^i\partial/\partial p_j) $ are coexisting with the non commutative theta form structure This requires consistency conditions ,for example , due to the Jacobi identity.It is noted that a derivation and a commutation get together in the Lie algebra of Poisson Brackets, leading to the Heisenberg equations for the quantum theory and the Heisenberg group algebra on the dynamical variables.There are now two commutation and derivation structures operating, due to the noncommutative theta form.

  \qquad 3.The Particle and Gauge Field system

In the presence of an interacting gauge field, the dynamical variables could be written with the congugate variables as denoted by $ c_i = (\theta^{-1})_{ij}x^j   + A_i$ and $\nabla_i f =\partial_i f -i(A_i * f- f*A_i)$; such that  a gauge symmetry acts  $ g*c_i*g^+$ and a covariant derivative $\nabla_if = \partial_if  - i[A_i,f  ]$ is defined. Including the noncommutative form gives the derivation as: $\nabla_i f =if*c_i -ic_i*f $ and the 'curvature' form of the gauge field as : $ F_{ij} =-i (c_i *c_j -c_j *c_i) +(\theta^{-1})_{ij} $.  The integrand of the action for the noncommutative Yang Mills field is : $ S_{YM} =-1/4g^2  Tr \sum (-i[c_i ,c_j] +(\theta^{-1})_{ij})^2 $; with $ i\neq j$ .

  The fields are defined with $ A_i -> gA_i g^+ -ig \partial_i g^+ and \qquad F_{ij}-> gF_{ij} g^+ $ . A derivation $ df =-[\theta,f];\qquad  d[\theta,f]=-d^2f=0 ;\qquad d\theta +\theta^2 =e $ is an identity. Where\qquad  $ g^+ =g^{-1} +i/2 \theta^{ij}g^{-1}(\partial_ig)g^{-1}(\partial_jg)g^{-1}$ . The fields are \qquad $ F_{ij}= \partial_iA_j -\partial_jA_i-i[A_i,A_j] +i/2 \theta ^{kl}[\partial_kA_i\partial_lA_j-\partial_kA_j\partial_lA_i]$ . In the usual notation the gauge connection form for the gauge group $ g $ , with group elements $T_a $and structure constants $f^{c}_{ab}$, for the non Abelian group is \qquad $  T_a A^a_i ;\qquad  [T_a ,T_b] = f^{c}_{ab}T_c ;\qquad   F =dA +A\wedge A $.

  \qquad  4.The Action Integral of the Particle and Gauge  Field System.

The Yang Mills action is\qquad  $ S_{ym} = -1/4g^2\int d^{D}x tr_N ( F_{ij} * F^{ij})$. Collecting all the terms of the various commutation, derivation structures, and keeping the relevant symmetries of the particle/field system ;the total action integral is written. This leads to the path integral partition function  $\int d[x,A ] exp(iS_{ym}) $ on any manifold $ M $ with genus $ h $ and Euler number  $ 2-2h.$ The construction with the Connes spectral triple(noncommutative algebra,Hilbert space(f) ,operators( F) on the space ; $df=i[F,f],F^2=0,d^2=0$) and index theorems is made.  The presence of the two form $\theta $ creates  a term in the index of operators , $ Tr\theta^2/96\pi^2 $ in addition to $ - Tr F^2/8\pi^2 $. This is expected to modify the gauge fixing and ghost terms in the action integral. There is a possibility that it would affect regularisation conditions. [Ref ].

 In a model of  hadrons in which the Yang -Mills field interacts with quantum dynamical variables of the constituents of the 'bag' model; a length scale exists and a noncommutative $\theta $ form could be defined.  The gluonic propagators would be regularised, and the quark ,gluon second quantised Fermionic and Bosonic operators would have a non commutative structure and Bogoliubov transforms . Does the 'bag' model length scale give new results about the quark-gluon excitations in a hadron, in a noncommutative non Abelian Yang -Mills QCD. That is still an open question.In the presence,also, of Fermi -Bose exchange operators , the combination of these various structures in the theory  ,could yield some interesting physics. The analogy is to the quantum phase space and noncommutative structure introduced in the Abelian electromagnetic field in interaction with fermions (electrons) in a quantum Hall effect type of Hamiltonian. The magnetic length provides a length scale, and interesting physics has developed even in the first quantised version.[Ref ].

  \qquad 5. Symmetries and Transforms

The Fermi- Bose symmetry is introduced : $ Q|F> =|B>  $ and $ Q^+|B> =|F> $
with $ [Q,Q^+] =1$. Then the Witten index is defined for the exchange as dim Ker$(Q-Q^+)$ given by $Tr( (-1)^Fexp(-\beta H))$ ; with $ H = Q^+Q $ , $ [Q,H] =0 $  . In the presence of a commuting and anticommuting number system ,generally $ ab= qba$ ; the commutator and anticommutator algebra, of the respectively, Bosonic and Fermionic dynamical variable operators can be unified . The developments in quantum groups give the structure of q deformed algebras. The Bogoliubov transforms on the corresponding $( a_i  , a^+_j)=\delta _{ij}$,are also defined in a common way as $ \alpha_{ik} a_k  +\beta_{ik} a^+_k  = b_i$ and$ \gamma_{ik}a_k  +\delta_{ik}a^+_k = b^+_i $ so as to preserve the algebra of the commutation structure. The conditions are  $\alpha \delta -\beta \gamma^T =I ;\alpha\gamma +\beta \delta^T =0 $. There are similar relations for the anticommutator and commutator algebras .

It is this property that permits the Fermi -Bose symmetry to be a starting point for supersymmetry. Similarly on the  q deformed commutation, the Bogoliubov like transforms can be defined, that preserve the algebra as automorphisms.  $ aa^+ +qa^+a =1 , bb^+ +qb^+b =1 ; \alpha \gamma +q\gamma\alpha =0 , \beta\delta +q\delta\beta =0 ;and  \qquad \alpha\delta +q\gamma\beta =1 ; \beta\gamma +q\delta\alpha =q $.These have to be compatible with the additional $\theta$ form structure, and the gauge symmetry.For example \qquad $[x^i,x^j]=i\theta^{ij} ;[p_i,p_j]=i \theta_{ij},$ and \qquad $a^i =1/\sqrt 2 (x^i +p_j) ;\qquad [a^i,a^j] = i\theta^{ij}-i\theta_{ij};\qquad [a^i,a^{j+}] =i\theta^{ij} -i\theta_{ij} +i\delta^{ij} $.

  \qquad  6. Partition Function Path Integrals  of the Physical system

A complete framework should utilise the possible transforms and their consistencies to manifest the symmetries in the particle -field system.In addition to the Yang Mills action there is a generalised quadratic form. The action integrand includes the terms $ \chi^+h\chi$ ,where  $\chi $ is the set of Bosonic/Fermionic or q deformed variables,in phase space of  $x^i$ and $ c^i$ canonically congugate variables. $ h$ is the quadratic form  on this space.The matter field couples to the gauge field ;with $ N/g^2 tr(\nabla\chi +ig[A,\chi])^2 $, The partition function can be written with the total action S, as $ \int d[\chi,A]exp(iS) = Z$,for the field and matter and their interaction.The introduction of the usual $J'\chi $and $JA$ terms in the action will give the functional derivatives with respect to J and J' of the Z[J,J'] functional. Such as $(\delta Z[J] /\delta J)Z[j]^-1$, and hence the 'n' point Green functions, and the relation between the free Z[J,J'] and the interacting Z[J,J'].This completes the framework of the theory.With the second quantised version of the particle variables $a^i$ and $ a^{+i}$, as the usual linear combinations of the congugate variables $x^i $ and $c^i$ ,or $\chi$ in phase space; the vacuum to vaccum amplitude $<0|0>$ can be explicitly calculated; subject to the Bogoliubov transform and other conditions.  

  \qquad  7.Consistency of the structures ,including Noncommutative Geometry

The first quantised dynamical variables  have the classical and quantum phase space correspondence in the Heisenberg , Wigner representation. The noncommutative * product of two functions is a generalisation of the Fourier transform ; $ \int d^{n} k d^{n} k' (2\pi )^{-2n} f(k)g(k'-k) exp(-i/2 \theta{ij}k_ik'_j exp(ik'_i x^i))$. Composition rules for Wigner functions , Weyl operators:  $ W[f] W[g] =W[f*g]$; $[\partial_i ,W[f]] =W[\partial_i f]$.  $ W[f] = \int d^n x f(x) \int d^{n} k (2 \pi)^{-n} exp(ik_i X^i) exp(-ik_ix^i)  $. In the reduced symmetry of the Lie algebra,the  Poisson Brackets  equal zero for coordinates and momenta among themselves and equal to one for the congugate variables.. The additional noncommutative structure , now introduces non zero commutation among the coordinates as well as the momenta among themselves. This could be seen as the realisation of the canonical and noncommutative structures, that make the interpretation of position like and congugate momentum like variables, indistinguishable .

 That is the full nature of the symplectic manifold , modulo non commutative geometry. For classical phase space and its quantum correspondence,besides the Wigner function and Heisenberg group representation, the $\theta$ two form gives another realisation of the Weyl group, Moyal bracket representation.This $\theta $ could be seen as  a two form on the tangent bundle, and its counterpart on the cotangent bundle. It is global, in the sense that it has all orders of derivatives . And it is locally consistent with the invariant symplectic two form; that is the Poisson/ Lie bracket structure of infinitesimal transformations.

  \qquad  8. Fermionic/Bosonic and q deformed system and the Theta structure

The additional $\theta$ form noncommutative structure introduces on the physically relevant dynamical variables , a two form which is symmetric for Fermionic and skew symmetric for Bosonic variables . This can be again seen as a special case of a commutation structure in a unified way, by considering both underlying number systems together, and the joint quantisation of commuting and anticommuting variables. Such a two form ,now induces additional conditions on the preservation of the $\alpha,\beta, \gamma,\delta $ matrices of the general linear canonical transforms of the Bogoliubov type.The more general 'q' deformed algebra is used ; $ a_ia^+_j +qa^+_ja_i =\delta_{ij} $, then the full quantum group transforms act on the quantum operators. There will be consistency requirements on the two structures of $\theta $ forms and the 'q' deformed one.

As a simple example consider $ [x^1,x^2] =i\theta^{l2}$ and $ x^i -> x^i +i\theta \epsilon ^{ij}A_j $ with $A_i =B/2 \epsilon_{ij} x^j$ so that the commutator varies as $ 1/B$. The operators with $ a =\sqrt{B/2} (x^1 + ix^2)$ is then the the choice for quantisation. There is a special case of a q deformation as ,$ exp(ix^1) exp(ix^2) = exp(-2\pi i\theta) exp(ix^2)exp(ix^1)$, where $\theta = (hc/2\pi eB)$,is a magnetic length parameter. A gauge covariant derivation and commutation, compatible with the $ \theta $  skew form: $\nabla_i\phi =\partial_i\phi +i[A_i,\phi]$. These are useful variables for Laughlin/Landau analysis in Quantum Hall effect. In a model for quantum geometry the skew form length scale is the Planck length. Area forms constructed from these variables could be compared with those in the quantum Riemann geometry, and with fuzzy geometry on the plane,and sphere. The fundamental 'displacement' variables are:   $x^i -i\theta^{ij}\partial /\partial x^j$ , and the generator is $ exp(-i \partial /\partial x^i \theta^{ij}\partial / \partial x^j)$.

  \qquad  9.Bogoliubov Transforms and Theta structure

In another example , the variables  have; $ [x^i,x^j]= i \theta^{ij} = i/B \epsilon^{ij}; [p_i,p_j] = i B \epsilon_{ij} ; \qquad[x^i , p_j]= i\delta^i_j ;\qquad 1/\sqrt 2(x^i +i\delta^{ij} p_j)=a^i $ and its adjoint; with \qquad $[a^i ,a^{+j}] =\delta^{ij}$. The Bogoliubov transforms are defined on these operators, with the consistency conditions.$ [a_i,a^+_j]= i/2( \theta ^{ij} -\theta_{ij}+\delta_{ij}$ ; becomes $i/2(\alpha \gamma +\beta\delta)(\theta -\theta^{-1})$ and $[a_i,a_j] =0$ becomes $(\alpha^2+\beta^2)(\theta-\theta^{-1}) +(\alpha\beta -\beta\alpha)(\theta-\theta^{-1}-\delta)$.And similarly for the adjoint and the $\gamma.\delta$ case. With the  q deformed algebra; these equations remain valid with the multiplier q on the right hand side.Hence all three conditions,namely $\theta$ form,  q deformed operators and Bogoliubov transforms can be made conditionally compatible. The operators $ c_i $ , instead of the momenta $ p_i$ ,which included the gauge field connection; could also be made consistent with these structures. General invariance under this extended symmetry would give rise to the physically permissible theories . The Fermionic/Bosonic particles interacting with Yang- Mills gauge fields is one of the fundamental developments of physics that is revised with the new symmetries and noncommutative geometry structures.

  \qquad  10. Conclusion

The theory of particles interacting with fields has generalised to include the noncommutative structures on the dynamical variables. The framework requires  consistency among the symplectic/Poisson structures of derivation /commutation and that of the $\theta $ form. The quadratic form on the particle phase space and the Bogoliubov like transforms of the operator algebra can be written for commutators and anticommutators and in the q deformed case. This can be made consistent with the noncommutative structure. The gauge field symmetries have a commutation and derivation ,which is also made compatible with the $\theta$ form. Action functionals describing the physics of the Fermionic/Bosonic particles interacting with the Yang Mills gauge fields could give a quark-gluon model, with  length scale parameter for the new $\theta$ form. Although some of the interrelations of the fundamental structures are being discovered ;more work is needed ,both on specific problems of applications and that of formulation from first principles of new physics.

Acknowledgements

 * The facilities, discussions and hospitality at the Institute of Mathematical Sciences,C.I.T campus, Tharamani, Chennai ,600113, India are gratefully acknowledged.

References

(1) J.Madore ,'Noncommutative Geometry for pedestrians' arxiv: gr-qc/9906059

(2)J.Madore 'An introduction to Noncommutative Differential Geometry and its physical applications' Cambridge University Press 2nd edn 1999.

(3)A.Connes 'A short survey of noncommutative geometry' Jnl of Math Phys vol 41,6, June 2000.

(4)M.Douglas ,N.Nekrasov 'Noncommutative field Theory 'Rev Mod Phys v73,2001,977-1029

(5)R.Szabo 'Quantum Field Theory on non commutative spaces'Physics Reports,v378,,4, May 2003.

(6)J.Madore, S.Schraml,P.Schupp,J.Wess 'Gauge theory on non commutative spaces'' arxiv:hep-th/0001203.

(7)M.Henneaux, C.Teitelboim, 'Quantisation  of Gauge systems' Princeton univ.press 1992.

(8)M.Reuter,'Noncommutative geometry on quantum phase space' hep-th/9510011

(9)R.Coquereaux,J.Geom.Phys 11(1993),307

(10)R.Coquereaux, 'Noncommutative geometry and Theoretical physics' J.Geom Phys6(1989)425.

(11)J.Frohlich, O.Grandjean,A.Recknagel,'Supersymmetric quantum theory and noncommutative geometry',math-ph/9807006.

(12)S.Majid,'Quantum groups and noncommutative geometry'

,hep-th/0006167

(13)R.Oeckl,'The quantum geometry of spin and statistics' hep-th/0008072.

(14)O.Alvarez,'Lectures on Quantum mechanics and the Index theorem',

I.A.S.Mathematics series,vol 1,1995.

(15)M.Blau,'Quantum Yang Mills theory on arbitrary surfaces'

(16)S.Cho,R.Hinterding,J.Madore ,H.Steinacker-'Finite field theory on non commutative geometry',hep-th/9903239

(17)T.R.Govindarajan,'Quantum gravity-Recent developments,Pramana, Indian Academy of sciences,2003

(18)N.Seiberg,E.Witten'String theory and noncommutative geometry',hep-th/9908142
(19)E.Langmann,'Quantum Gauge theories and non commutative geometry',hep-th/9608003.

(20)Y.I.Mannin'Quantum Groups and noncommutative geometry 1988

(21) hep-ph/ 9307209,hep-th/9604146,hep-th/9605001,hep-th/9311179,hep-th/0005210,hep-th 0206007,hep-th9701078,physics/9709045.

(22)F.Iachello,P vanSecker, 'The interacting boson-fermion model-nuclear spectra.Cambridge University press 1991.

(23)A.Connes,'Noncommutative differential geometry'IHESPublv62(1986),257

(24)J.Pati,'Bose-Fermi symmetry',hep-ph/9506211

(25)V.Nair,'Noncommutative gravity',Talk at the Noncommutative geometry and field theory workshop at Institute of Mathematical sciences 2003,imsc.res.in

(26)A.Dhar 'Loop equations in Noncommutative geometry field theory',Talk at the IMSC workshop 2003.

(27)'Noncommutative geometry and quantum Hall effect' cond-matt/9411052

(28)A.Ashtekar,A.Corichi,J.Zapata,'Quantum theory of geometryIII:Non commutativity of Riemannian structures.',Class.and quant.grav.v15(1998),2955.

(29)D.Kastler ,'Noncommutative geometry and fundamental physical interactions',Lecture notes in Physics,Springer Verlag,Proceedings(1999)

(30)J.Lewandowski-'Constraints on unified gauge theory fromNon commutative geometry' Mod Phys Lett,A,vol11,1996,2561

(31)J.Isidro,'The geometry of Quantum mechanics'-hep-th/0110151

(32)M.Nakahara,Geometry,Topology and Physics,Adam Hilger 1990

(33)A.Polyakov, Gauge fields and strings,Harwood 1987

(34)T,Dass, 'Symmetries,Gauge fields,strings and fundamental interactions',Wiley Eastern 1993.

(35)V.Nair ,A.Polychronakos,'Quantum mechanics on the non commutative plane and sphere',hep-th/0011172

(36)J.Leinaas, J.Myrheim, Phys Rev ,B37 (1988),9286

(37)I.Pris, T.Schucker,'Noncommutative geometry beyond the standard model',J.Math Phys,38(1997),2255

(38)G.Landi-'An introduction to noncommutative spaces and their geometries'.Lecture notes in Physics-New series M,Monograph 51,springer Verlag, Heidelberg(1997)

(39)J.Frohlich,O.Grandjean,A.Recknagel,'Supersymmetric quantum theory, noncommutative geometry and gravitation'Lecture notes,Les Houches 1995,hep-th/9706132

(39)A.Connes, J.Lott 'Particle models and noncommutative geometry' ,Nucl phys(proc.suppl),B18(1990),29

(40)A.Connes,'Noncommutative geometry and reality',Jnl Math Phys,

36(1995),6194

(41)A.Connes'Action functional in noncommutative geometry' Comm math phys 117(1988),673

(42)A.Connes ,'Gravity coupled with matter and foundation of noncommutative geometry, Comm Math Phys 192(1996),155

(43)A.Chamseddine,A.Connes ,'Universal formula for noncommutative geometry action:Unification of gravity and standard model',Phys Rev.Lett(1996),4868

(44)A.Chamseddine,G.Felder,J.Frohlich,'Gravity in noncommutative geometry',Comm Math Phys 155(1993),205

(45)Esposito,'Spectral Geometry'.

(46)A.Ashtekar,'Quantum geometry and gravity:Recent advances',gr-qc/0112038

\end{document}